\documentclass{emulateapj}

\newcommand{\etal}{{et al.}}
\newcommand{\eg}{{\it e.g.,}}
\newcommand{\ie}{{\it i.e.,}}

\begin{document}

\title{A Disk Galaxy of Old Stars at z\,$\sim$\,2.5\footnotemark[1]}

\footnotetext[1]{Based in part on data collected 
at Subaru Telescope, which is operated by the National Astronomical Observatory of Japan. 
Some of the data presented herein were obtained at the W.M. Keck Observatory, which is 
operated as a scientific partnership among the California Institute of Technology, the 
University of California and the National Aeronautics and Space Administration. The 
Observatory was made possible by the generous financial support of the W.M. Keck 
Foundation.}

\author{Alan Stockton}
\affil{Institute for Astronomy, University of Hawaii, 2680 Woodlawn
 Drive, Honolulu, HI 96822}

\author{Gabriela Canalizo}
\affil{Institute of Geophysics and Planetary Physics and
Department of Earth Sciences, University of California, Riverside, CA 95521}

\author{Toshinori Maihara}
\affil{Department of Astronomy, Kyoto University, Kitashirakawa-Oiwake-cho,  Sakyo-ku, Kyoto 606-8502, Japan}

\begin{abstract}
We describe observations of a galaxy in the field of the $z=2.483$ radio galaxy 4C\,23.56, 
photometrically selected to have a spectral-energy distribution consistent with an old stellar
population at the redshift of the radio galaxy. Exploration of redshift---stellar-population---reddening
constraints from the photometry indicates that the galaxy is indeed at a redshift close to that
of 4C\,23.56, that the age of the most recent significant star formation is roughly $\gtrsim2$ Gyr, 
and that reddening is fairly modest,
with more reddening required for the younger end of stellar age range.  
From analysis of a deep adaptive-optics image of the galaxy,
we find that an $r^{1/4}$-law profile, common for local spheroidal galaxies, can be excluded
quite strongly.  On the other hand, a pure exponential profile fits remarkably well, while the best fit is
given by a S\'{e}rsic profile with index $n=1.49$. Reconstruction of the two-dimensional form of the
galaxy from the best-fit model is consistent with a disk galaxy
with neither a significant bulge component nor gross azimuthal structure. The
assembly of roughly $2L^*$ of old stars into such a configuration this early in the history of the universe is not easily explainable by any of the currently popular scenarios for galaxy formation. A
galaxy with these properties would seem to require smooth but rapid infall of the large mass of 
gas involved, followed by a burst of extremely vigorous and efficient star formation in the resulting disk.
\end{abstract}


\section{Introduction}

Our understanding of the mechanisms and timescales of the formation of the very first
luminous galaxies in the universe is still very uncertain. 
Many detailed studies of local luminous early-type
galaxies favor formation at very early epochs and 
within a very short time span (\eg\
\citealt{pee02} and references therein).  In particular, the 
$\alpha$ elements are enhanced in luminous ellipticals relative to Fe (\eg\ \citealt{wor92}), this enhancement is strongly correlated with the central
velocity dispersion $\sigma$ \citep{tra00}, and the relation is 
similar for field and cluster ellipticals \citep{ber98,col99,tho02}. 
Galaxies with the oldest stellar populations are both the most luminous and the
most $\alpha$ enhanced \citep{tho02}.  If these claims are valid, the 
most luminous galaxies formed within a quite short
period very early in the history of the universe, less luminous early-type galaxies
formed more gradually and somewhat later, and differences in star formation histories
between ellipticals in the field and those in clusters are relatively minor. This last
statement is also supported by fundamental-plane analyses (\eg\ \citealt{vanD03, treu01}; but
see also \citealt{treu02}).  Such conclusions appear to be
difficult to reconcile with standard interpretations of the results of the semi-analytic 
cold-dark-matter (CDM) models (\eg\ \citealt{tho99,kau00}), which predict that, for the currently favored
low-density cosmology ($H_0\sim70$, $\Omega_m\sim0.3$, $\Omega_{vac}\sim0.7$;
assumed throughout this paper),
most of the present-day bright ellipticals will have undergone final assembly 
between $z=1$ and $z=2$. However, there are at least two caveats to this statement:
(1) One of the least certain parts of the semi-analytic CDM models is the treatment of star formation
and resulting feedback effects, and it is possible that different assumptions in this area could
mitigate at least some of the discrepancy; and (2) {\it assembly} at relatively low redshifts
is not necessarily inconsistent with formation of most of the stars at much higher redshifts.
Nevertheless, both the color-luminosity relationship for spheroids \citep{bow92,ell97,vanD00}
and the correlation of
$\alpha$-element enhancement with luminosity show that the formation of the
stellar population and the assembly of the galaxy cannot be completely 
independent processes. As \citet{pea99} has succinctly put it, ``It seems as if the
stars in ellipticals were formed at a time when the depth of the potential
well that they would eventually inhabit was already determined.''

In this context, it is of considerable interest to seek out and to investigate the galaxies
with the oldest stellar populations at any given redshift. While such galaxies may not be
typical, they can establish one end of the range of realistic formation scenarios, and they
likely have less complicated star-formation histories than most. Furthermore, they
may, in fact, be more representative of the formation of the bulk of the stars in
early-type galaxies generally than might at first seem likely, since
a miniscule contamination of young stars in a galaxy can drastically alter its rest-frame ultraviolet 
colors and eliminate such galaxies from color-selected samples.

There has been much recent activity in observations and discussion of so-called
{\it extremely red objects} (EROs; \eg\ \citealt{liu00,dad00,dad03,sto01,cim03,yan03,
miy03,gil03}; earlier references can be found in these papers).  These objects, which 
are often defined to have $R\!-\!K\gtrsim6$ or $I\!-\!K>4$, include
two disparate classes.  Some are highly reddened high-redshift starbursts 
(\eg\ \citealt{dey99}).  The remainder are almost exclusively galaxies
with $z\gtrsim1.2$ whose stellar populations are already
ancient and have little admixture from more recent star formation \citep{sto95,
dun96,spi97,liu00,soi99,sto01}.
For convenience, we shall 
henceforth simply refer to this latter class as ``old galaxies'' (OGs); but note once again that
this term refers to the age of the stellar population, with no necessary
implication regarding the age of assembly of the galaxy itself, aside from the sorts of
constraints mentioned above. 
In any case, these OGs seem to represent the earliest major episodes of star
formation in the universe.

The actual ages of the dominant stellar population in the small number of
identified OGs that have been studied in detail have generated some
controversy because of
disagreements about spectral synthesis models and their interpretation
(see, \eg\ \citealt{dun99} and references therein).  However, a minimum age
of 2 Gyr for the oldest objects at $z\sim1.5$ is accepted by virtually everyone, and
there are good reasons for believing that some, at least, have ages between
3 and 4 Gyr \citep{dun99,sto01,nol01,nol02}. If these
latter age estimates are correct, then the precursors of these galaxies will still
have essentially fully formed stellar populations at higher redshifts.  For example,
for our assumed cosmological parameters, galaxies at $z=2.5$ will be $\sim1.6$
Gyr younger than at $z=1.5$. This means that $z=1.5$ galaxies for which the last major 
episode of star formation occurred between 3 and 4 Gyr earlier might be expected
to be fairly well settled and have small average internal extinction at $z=2.5$, and
probably at even higher redshifts.

Many surveys of EROs have made the assumption, explicit or not, that a dominant
spheroidal component (in mass, if not in light) is a necessary condition for the presence
of an old stellar population at high redshifts \citep[\eg][]{mor00,smi02b}.
In this paper, we describe the first object among those found in our search for OGs in radio 
source fields at $z\sim2.5$ that we have been able to investigate in some detail, and we
show that it is a luminous disk galaxy, with little or no bulge component, comprising
very old stars.

\section{Selecting Galaxies with Old Stellar Populations in Radio-Source Fields at 
$z\sim2.5$}\label{photsect}

We have recently concluded a search for OGs in the fields of quasars at
$z\sim1.5$ \citep[][Stockton \etal, in preparation]{sto01}. Radio source fields are 
generally likely to include regions of higher-than-average
density in the early universe (\eg\ \citealt{bes03}), in which processes of galaxy formation and evolution
are expected to have proceeded more rapidly than in the field. More importantly, looking for
companions to radio sources at a specific redshift allows us to choose redshifts for
which the photometric diagnostics from standard broadband filters give the cleanest separation
between old galaxy populations and other possible contaminants, such as the dusty
starbursts mentioned above. The relevant key redshifts for our present purposes
are $z\sim1.5$ (4000 \AA\ break just shortward of the $J$ band) and $z\sim2.5$ (break
between $J$ and $H$ bands).

Assume that, for the brighter OGs from our sample at $z\sim1.5$, essentially all of
their stars formed within 500 Myr of the Big Bang (corresponding to a stellar population age of
3.7 Gyr at $z=1.5$).  For passive evolution of the stellar population, one finds from solar
metallicity models (see \citealt{bru93, bru03}; we have used the 2000 versions of the
models for analysis here unless otherwise specified) that the
precursors of such galaxies at $z\sim2.5$ would have $J\sim22.4$, $H\sim20.5$, and $K'\sim19.5$.
This assumption is deliberately an extreme one: if our assumed age (corresponding to $z_{formation}\sim9$) is too high, then these magnitudes are all a bit pessimistic.
Our approach has been first to obtain imaging to $K'\sim20$ ($10 \sigma$) and 
$J\sim23$ ($5 \sigma$), looking for objects in the field with $J\!-\!K'\sim3$. If any
likely candidates are present, we obtain deeper imaging at $J$ and $K'$, and include imaging
in the $H$ band. Most of the infrared observations (including all of those relevant to this
paper) have been obtained with the infrared camera CISCO
\citep{mot02} on the 8.2 m Subaru telescope on Mauna Kea. The detector is a Rockwell 
$1024\times1024$ Hg-Cd-Te Hawaii-2 array, with a pixel scale of 0\farcs105 and thus 
a field of just under 2\arcmin. Additional $R$-band imaging was obtained with ESI on the
Keck II telescope.

\begin{figure}[!t]
\epsscale{0.9}
\plotone{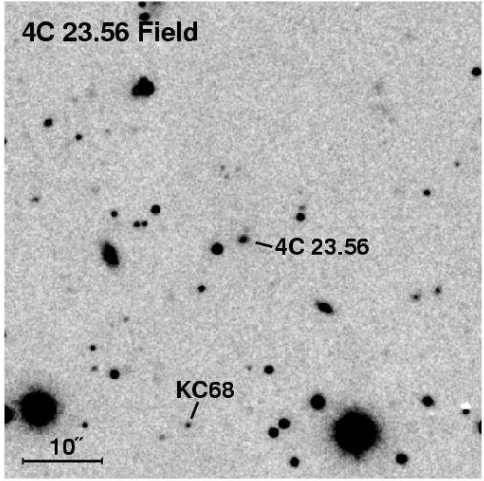}
\caption{Field of the radio galaxy 4C\,23.56. The image was obtained
with the CISCO IR camera on the Subaru telescope in the $K'$ band. The galaxy with a SED
indicating an old stellar population is indicated as KC68. The bright star to the lower-right side of the
image was used as the AO guide star.\label{fchart}}
\end{figure}

Our sample comprises 87 radio-source fields from the Texas catalog \citep{dou96} selected with the
help of the NASA/IPAC Extragalactic Database (NED) to have redshifts in the range
$2.3<z<2.7$ and $A_B<1$, based on the Galactic extinction estimates of \citet{sch98}. 
The coverage we have been able to achieve so far on these fields has been largely
determined by telescope scheduling and weather. We currently have adequate observations for
10 fields, of which at least 3 have objects that have well characterized spectral-energy distributions 
(SEDs) that closely
match those expected for old stellar populations at the redshifts of the radio sources, with
little intrinsic reddening. General results for this survey will be given elsewhere; here
we concentrate on one galaxy, found in the field of the radio galaxy 4C\,23.56, for which we
have both strong constraints on the SED and adaptive-optics (AO) imaging. A finding chart
for the object is shown in Fig.~\ref{fchart}. 
Although it was found independently in our survey, this galaxy was previously noted as a 
very red galaxy by \citet{kno96} in their 
comprehensive study of the field of 4C\,23.56. The galaxy is number 68 in their Table 4, and
we will refer to it as 4C\,23.56 KC68, or, more briefly, simply as KC68.


\begin{figure}[!tb]
\epsscale{1.0}
\plotone{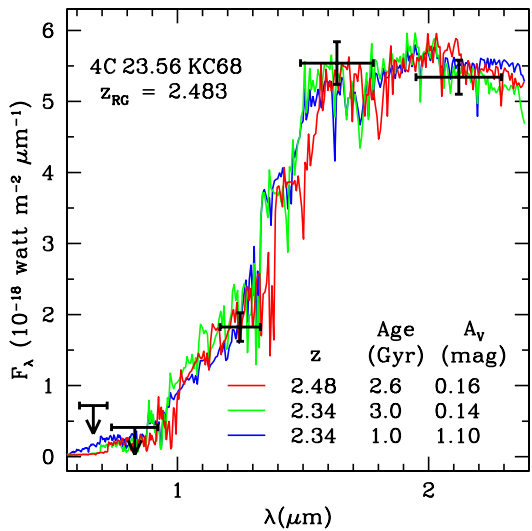}
\caption{Spectral-energy distribution in the observed frame for 4C\,23.56 KC68. The photometric bands are $R$, $I$, $J$, $H$, and $K'$; the $I$-band upper limit is from \citet{kno96}. The vertical 
bars show $1 \sigma$ errors, and the horizontal bars indicate the FWHM of
the filter. The upper limits shown are $3 \sigma$.  All of the photometric points and upper limits
have been corrected for Galactic extinction using the estimate of \citet{sch98}. 
The colored traces are examples of best-fit
redshifts, stellar populations, and extinctions, as discussed in the text.
\label{sed}}
\end{figure}
\begin{figure}[!tb]
\epsscale{1.0}
\plotone{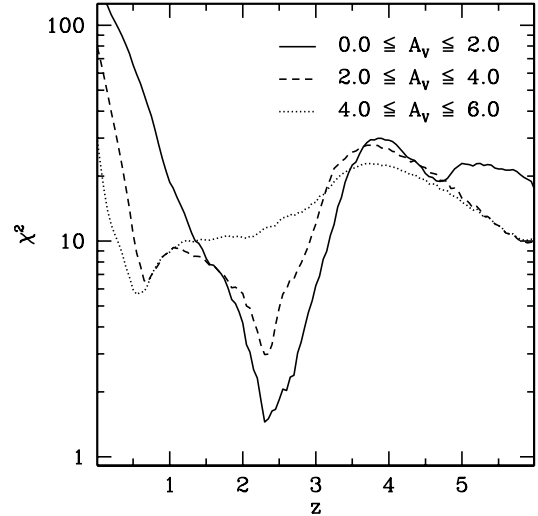}
\caption{Run of $\chi^2$ with $z$ for model fits to the photometry for 4C\,23.56 KC68. Models
with different ranges of internal extinction have been separated. At each redshift, the $\chi^2$
statistic for the best-fitting model has been plotted, so the ages and internal extinction vary along
each trace.\label{chi2}}
\end{figure}


The photometry for 4C\,23.56 KC68 is shown in Fig.~\ref{sed}. The magnitudes, prior to Galactic
reddening correction, are $R>26.7$ ($3\sigma$), $I>26.5$ ($3\sigma$), $J=23.38\pm0.12$,
$H=20.91\pm0.06$, $K'=19.90\pm0.03$.
We have used the photometric redshift code Hyperz\footnotemark[2] \citep{bol00} to explore 
\footnotetext[2]{\tt http://webast.ast.obs-mip.fr/hyperz/}
possible fits to the photometric data, and some of these model SEDs are also shown in Fig.~\ref{sed}.
The fitting procedure allowed redshifts $0\le z\le6$, 
internal extinction corrections $0\le A_V\le6$ (assuming a \citealt{cal00} reddening law), and
a full range of solar-metallicity stellar population models, including both single-burst models as well
as models incorporating various rates of continuing star formation. Within this parameter space,
several general conclusions emerged:
\begin{enumerate}
\item Only single-burst models produced good fits. This is presumably because significant continuing
star formation would weaken the sharp inflection shortward of the $H$-band point and/or violate the
$R$ and $I$-band upper limits.
\item The redshift is strongly constrained to the range $2.1\le z\le2.7$. Figure \ref{chi2} shows the
$\chi^2$ statistic for the fits as a function of redshift.
\item When redshift was left as a free parameter, the fits tended to home in on $\sim2.35$, with 
a 99\%\
confidence interval $2.1\lesssim z\lesssim2.7$. If one assumes the galaxy to be a member of 
a cluster associated with 4C\,23.56 and
therefore fixes the redshift at 2.483, a fairly respectable solution exists with an age of
2.6 Gyr and a modest internal extinction $A_V=0.16$ mag.
\item Although the lowest reduced $\chi^2$ was obtained for an old (3.0 Gyr) population with 
$A_V\sim0.14$, to some extent it is possible to trade younger populations for increased reddening.
However, this exchange cannot be carried too far: for ages $\lesssim1$ Gyr the rapid rise in the
flux at rest-frame 2000 \AA, coupled with the bluer colors longward of 4000 \AA, prevents the underlying
continuum from being massaged to give a reasonable fit to the photometry by any amount of 
Calzetti-like extinction.
\end{enumerate}
Table \ref{models} gives parameters for the three models shown in Fig.~\ref{sed}, 
which are representative of the range of plausible fits to the photometry.


\begin{center}
\begin{deluxetable}{ccccc}
\tablewidth{0pt}
\tablecaption{Model Fits to Photometry of 4C\,23.56 KC68 \label{models}}
\tablehead{\colhead{Model} & \colhead{$z$} & \colhead{Age\tablenotemark{a}} & \colhead{$A_V$} 
& \colhead{$\chi^2$}\\
\colhead{} & \colhead{} & \colhead{(Gyr)} & \colhead{(mag)}& \colhead{} }
\startdata
1 & 2.340 & 3.0 & 0.14 & 1.128 \\
2 & 2.483 & 2.6 & 0.16 & 1.832 \\
3 & 2.340 & 1.0 & 1.10 & 1.877 \\
\enddata
\tablenotetext{a}{Assuming solar metallicity. See text for further discussion.}
\end{deluxetable}
\end{center}


Note that, because of the well-known age-metallicity degeneracy, we can obtain younger ages
by allowing super-solar metallicities. Such a trade-off means, for example, that even though
some of the ages in Table \ref{models} exceed the age of the universe for the given redshift
and our assumed cosmological parameters (\ie\ 2.75 Gyr at $z=2.34$; 2.59 Gyr at $z=2.48$),
these excess ages can be considered proxies for high metallicities. At a redshift of 2.48, an 
$L^*$ elliptical with a 2.1-Gyr-old solar-metallicity stellar population would have $K'\sim20.5$ 
in the observed frame, so the observed $K'=19.8$ (after correction for Galactic extinction) for KC68
implies a luminosity of about $2L^*$. This mass associated with this luminosity is not strongly 
dependent on metallicity, the differential effect between solar and twice solar metallicities being
a reduction in the inferred mass by about 10\%.

\section{Adaptive-Optics Imaging of 4C\,23.56 KC68}\label{aosect}

\begin{figure*}[!tb]
\epsscale{0.9}
\plotone{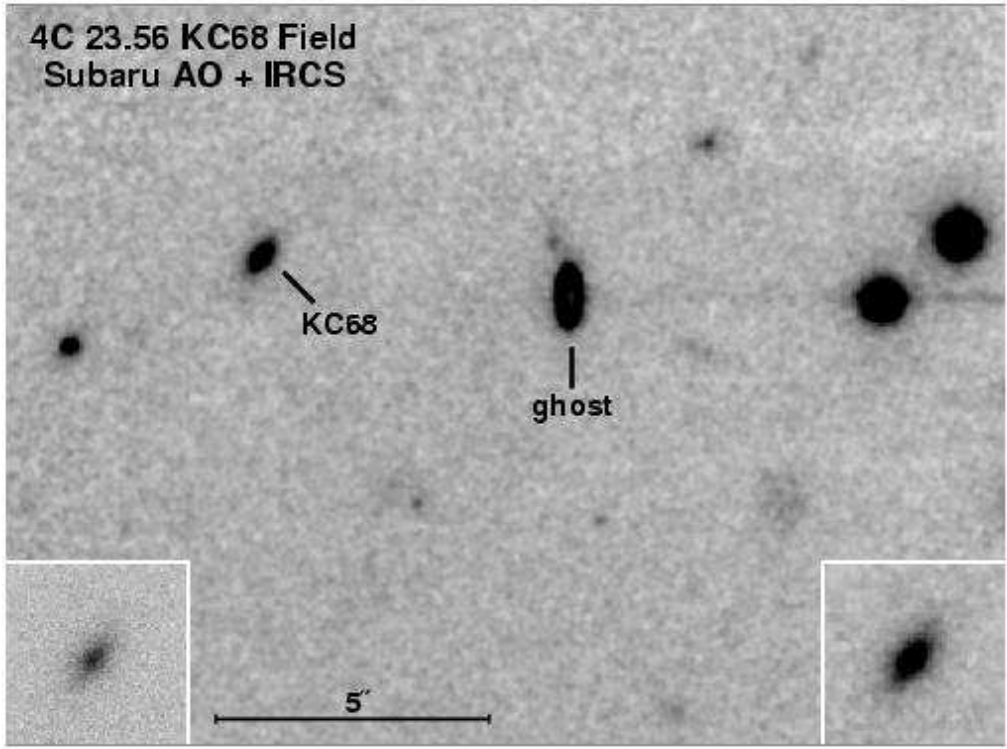}
\caption{Image of the field around 4C\,23.56 KC68 obtained with the AO system and IRCS on
the Subaru telescope. The main image has been smoothed with a Gaussian kernel with $\sigma=2$
pixels. The insets show KC68 enlarged by a factor of 2: the lower-left inset shows
the unsmoothed image at low contrast, and the lower-right inset shows an image at medium
contrast, smoothed with a Gaussian kernel with $\sigma=1$ pixel. A ghost image from the AO
guide star is indicated.}\label{aowide}
\end{figure*}

The galaxy 4C\,23.56 KC68 happens to lie $\sim20\arcsec$ from a 14th mag star, and we
have been able to obtain adaptive-optics (AO) imaging, using the Subaru 36-element,
curvature-sensing AO system and the
Infrared Camera and Spectrograph (IRCS; \citealt{kob00}). The $1024\times1024$ Aladdin3
array had a pixel scale of 0\farcs0225, giving a field of 23\farcs0. The total exposure was
12,600~s, all in the $K'$ filter. Figure \ref{aowide} shows a section of the AO-imaged field. The 
image cores have FWHM = 0\farcs10, and there is no evidence for significant anisoplanicity 
effects across the relevant part of the field. Insets in Fig.~\ref{aowide} show larger-scale
images of KC68 at two contrast levels. The galaxy clearly has a high degree of
symmetry, showing that $\sim2L^*$ galaxies, comprising relaxed
old stellar populations, can be in place by the time the universe is only 2.6 Gyr old.


\begin{figure*}[p]
\epsscale{0.9}
\plotone{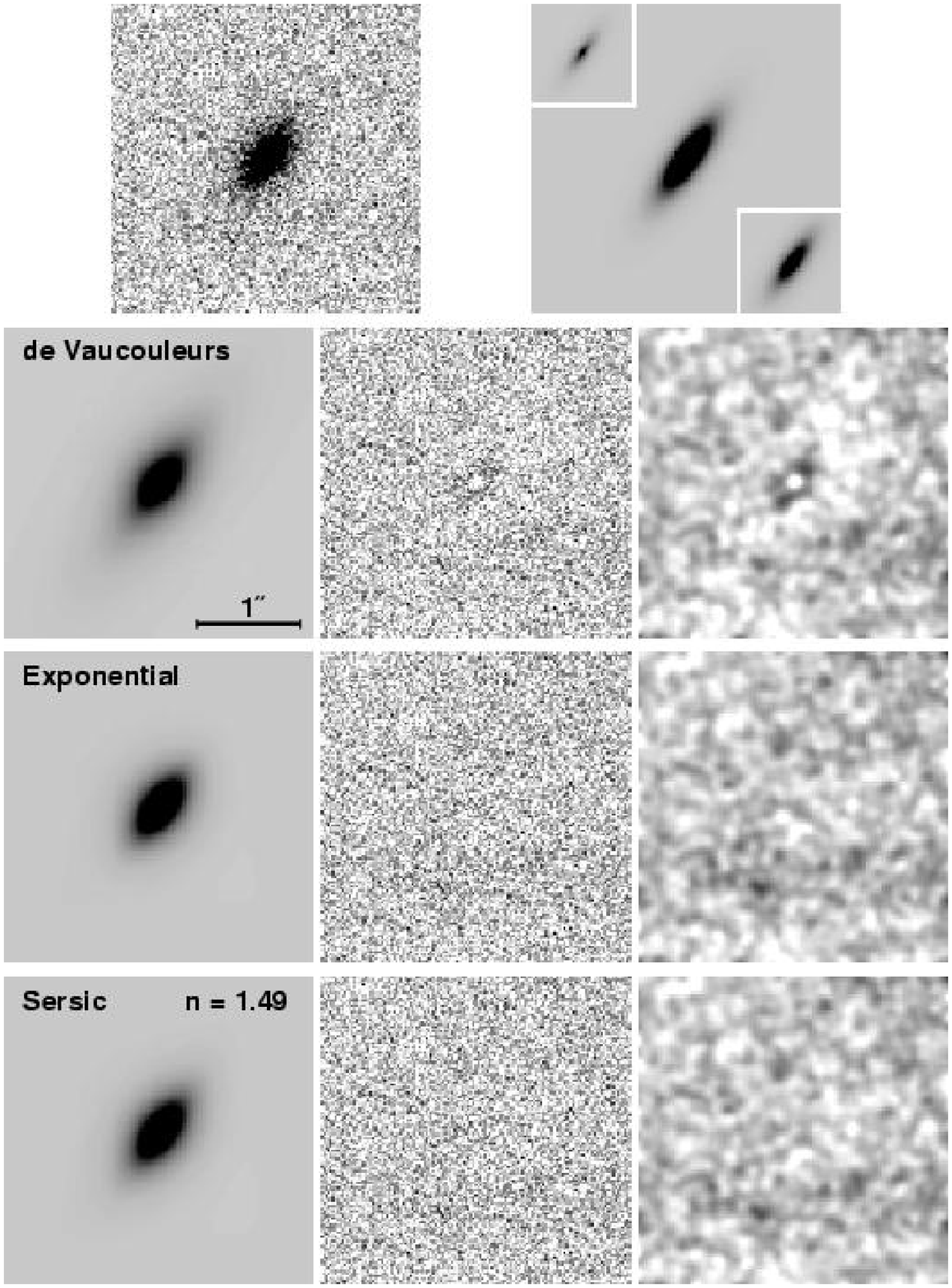}
\caption{\scriptsize Model fits to 4C\,23.56 KC68. The top-left image reproduces the observed image
from Fig.~\ref{aowide}. The three rows underneath show PSF-convolved
de Vaucouleurs, exponential, and
S\'{e}rsic models calculated with GALFIT  \citep{pen02}. The first panel of each row shows
the model, using the same stretch as for the observed image; the second panel shows the
residual after subtraction of the model from the image, again at the same contrast; the
third panel shows the residual after Gaussian smoothing with a $\sigma=2$ pixel kernel
and a factor of 3 contrast enhancement. Note that the de Vaucouleurs model systematically
oversubtracts at both small and large radii while undersubtracting at intermediate radii.
Differences between the exponential and S\'{e}rsic profiles are quite subtle, as expected from
the small S\'{e}rsic index. The top-right panel shows the appearance of the S\'{e}rsic model
before convolution with the PSF, which is presumably a good indication of the gross 
morphology of the galaxy.\label{2dmod}}
\end{figure*}

We can explore this morphology in more detail by making model fits to the image. We have used
C.~Y.~Peng's GALFIT\footnotemark[3] routine \citep{pen02} to fit two-dimensional models corresponding
to de Vaucouleurs ($r^{1/4}$-law), exponential, and general S\'{e}rsic  profiles\citep{ser68,cao93}, all convolved with the
PSF. Figure \ref{2dmod} shows these models and the residuals after subtraction from the observed
image. Although a pure exponential model is an excellent fit, the formal best fit is for 
a S\'{e}rsic model with an 
index $n=1.49$, an effective radius $r_e=0\farcs22$ ($=1.8$ kpc), and an axial ratio $b/a=0.33$. 
These numbers tend to confirm one's visual impression that
the galaxy is disk-like in morphology. While one cannot exclude the possibility of a small 
$r^{1/4}$-law bulge component, all of our attempts to model the galaxy as a two-component
($r^{1/4}$-law + exponential) system resulted in the $r^{1/4}$-law-component being given an
extremely large effective radius (regardless of the initial parameter settings), essentially smearing 
it into the sky background.
\footnotetext[3]{\tt http://zwicky.as.arizona.edu/$\tt\!\sim$cyp/work/galfit/galfit.html/}


\begin{figure*}[!t]
\epsscale{1.0}
\plottwo{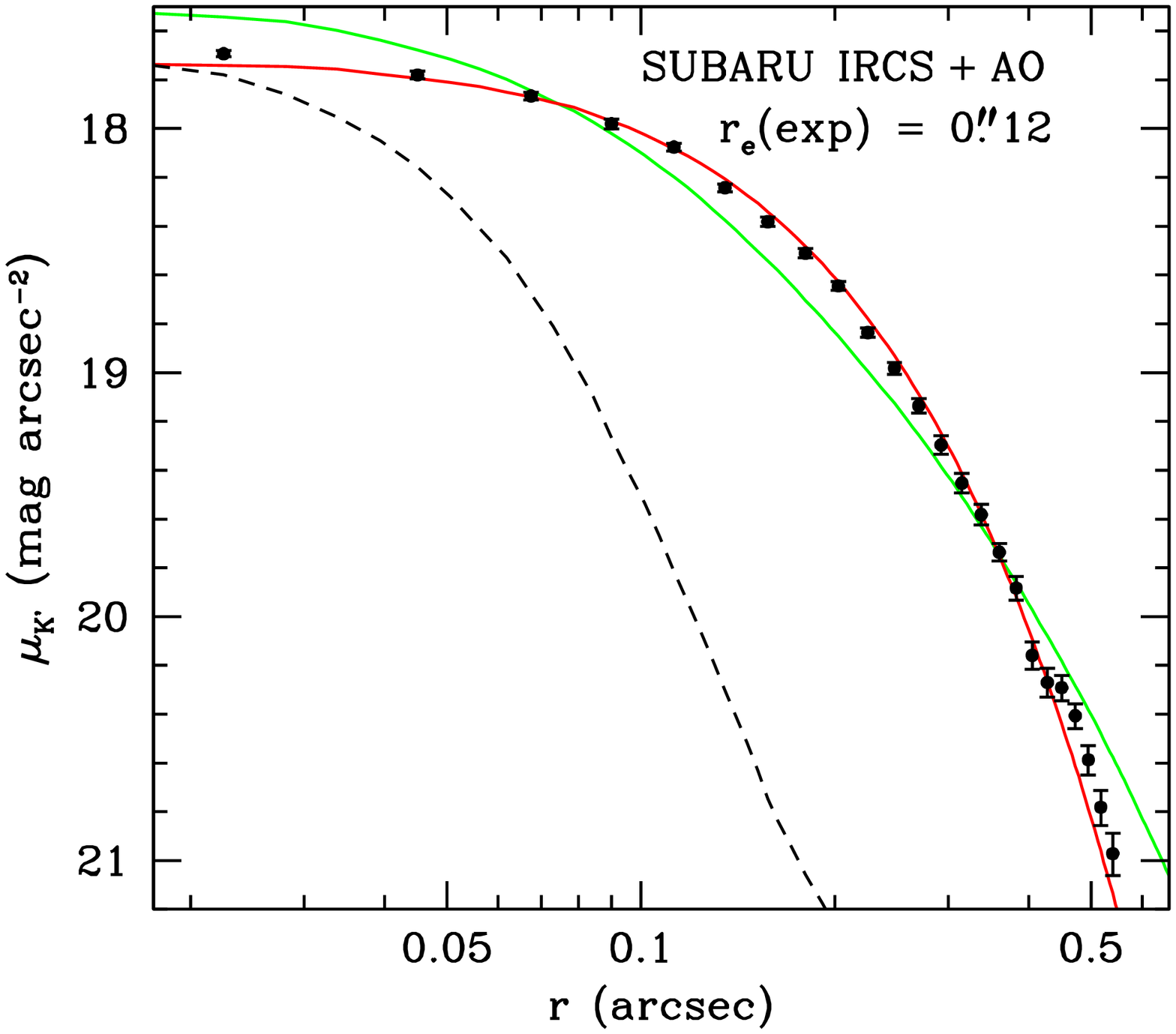}{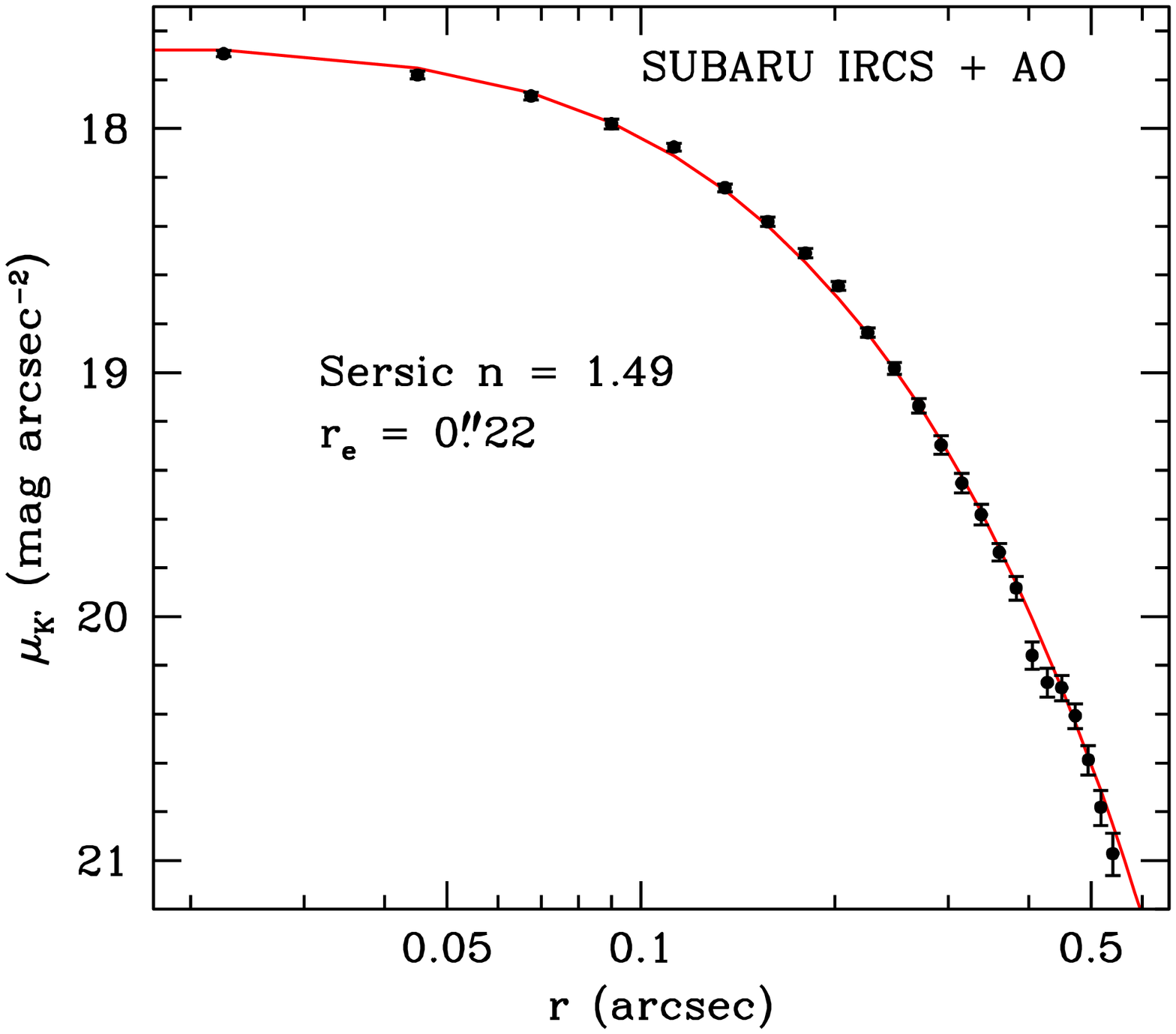}
\caption{The observed radial surface-brightness profile of 4C\,23.56 KC68 is shown
as the black points with error bars in both panels. In the left panel are plotted best-fit
de Vaucouleurs (green) and exponential (red) profiles from two-dimensional models
convolved with the PSF, along with the PSF
itself (black dashed line). The exponential model is clearly
a much better fit; the galaxy model itself (\ie\ before convolution) has an effective
radius $r_e=0\farcs12$ and an axial ratio $b/a=0.35$. The right panel shows the best S\'{e}rsic
profile fit, again after convolution with the PSF. The model has an index $n=1.49$,
an effective radius $r_e=0\farcs22$, and an axial ratio $b/a=0.33$.\label{sbp}}
\end{figure*}

For a more quantitative view, Fig.~\ref{sbp} shows the surface-brightness profile of KC68, along with 
seeing-convolved model 
fits. It is clear from the left panel that an $r^{1/4}$-law model cannot adequately fit the profile, 
and that an exponential model gives a much better fit. The right panel shows the best
S\'{e}rsic-profile fit.


\section{Summary and Discussion}
The identification of this galaxy and the other two (mentioned in \S\ref{photsect}) at $z\sim2.5$ that 
we have found 
so far with strikingly dominant old stellar populations demonstrates that at least some 
massive galaxies have completed virtually all of their star formation at sufficiently early times
that they are no longer heavily dust-enshrouded at this redshift.  Of these, at least 4C\,23.56 KC68 
has also achieved a relaxed regular morphology. The surprise is that this relaxed configuration 
seems to be very close to a pure disk, rather than to a spheroid, or even to a spheroid + disk.

There is abundant evidence that the large ($\sim40$\%) fraction of S0 galaxies seen in rich
clusters at the present epoch drops dramatically at higher redshifts \citep{dre97,fas00}, with the
slack being taken up by spirals, and with the proportion of elliptical galaxies remaining roughly
constant. The mechanisms for the apparent conversion of spirals into S0 galaxies have
been widely discussed ever since \citet{but78} first noticed the increasing fraction of blue
galaxies in clusters at high redshifts (see, \eg\ \citealt{bic02} for a recent example).

However, there is also evidence to support the classical view that at least a significant
fraction of the S0 population 
consists of galaxies that formed coevally with the oldest ellipticals \citep{tin76,ell97}.
\citet{iye00,iye03} have described an S0-like galaxy at $z\sim1.5$. In addition, recent large 
surveys find that a substantial fraction of objects identified as EROs have disks (\citealt{smi02a,yan03,gil03}; see also \citealt{smi02b}), and the
only galaxy in our own sample of
galaxies at $z\sim1.5$ with old stellar populations \citep{sto01} for which we have good AO 
imaging also is dominated by an exponential profile (Stockton \etal, in preparation). At a higher redshifts, 
\citet{lab03} have identified a sample of large, disk-like galaxies, one of which has fairly
red colors \citep{fra03} and is at $z=2.94$.

All of these cases appear still to have significant (and usually dominant) bulge components.
Indeed, one of the principal points of \citet{smi02b} is that the galaxy they describe may be a spheroid
in the process of acquiring a disk, validating a commonly proposed scenario for the evolution of
early-type spirals. In other words, the reason that this particular galaxy is classified as an ERO lies in the dominance of the bulge population and the weakness and low star-formation rate of the disk
component. Many other galaxies classified as EROs also show discrete star-forming regions, but
in these cases, too, the bulge population seems to dominate the colors used for the classification.

The galaxy we are discussing here poses a much more radical challenge to our 
understanding of the formation of old stellar populations at high redshifts: KC68 appears to be a 
$\sim2L^*$ galaxy of old stars having a disk-like morphology, with no significant bulge component. 
As indicated below, this galaxy seems so difficult to fit into the standard 
cold-dark-matter (CDM) picture of early galaxy formation that it is worth briefly
revisiting the steps by which we have reached this conclusion.

{\it Could the redshift be seriously in error?}---Certainly for low extinction, the photometric
redshift seems quite unambiguous (see Fig.~\ref{chi2}). For the \citet{cal00} reddening law, this
conclusion remains for the highest extinctions we have considered. We cannot exclude the possibility that one could contrive a reddening law that could result in a reasonable fit at a different redshift, for 
sufficiently high extinction. However,
the high degree of symmetry shown by the galaxy does not fit the fairly chaotic morphology usually
shown by galaxies having large amounts of dust. We believe that a redshift outside of the
range $2.1\le z\le2.7$ is extremely unlikely.

{\it Are the stars really old?}---We have seen in \S\ref{photsect} that reddening alone cannot
reproduce the observed SED, including the sharp inflection shortward of the $H$ band and
the tight constraints on the flux at rest-frame 2000 \AA\ and 2400 \AA\  given by the $R$ and
$I$-band upper limits. (This conclusion strictly assumes a \citealt{cal00} reddening
law; but the statement almost certainly holds for any plausible reddening law). It is necessary to 
have a strong break in the underlying spectrum, even if reddening is
substantial. Younger ages produce progressively worse fits, and ages $<1$ Gyr
do not seem at all likely, provided only that the stellar initial mass function (IMF) is not
pathologically deficient in high-mass stars. Of course, one can make the stellar population as
young as one wishes by an {\it ad hoc} truncation of the IMF above about $1.8 M_{\odot}$
(\eg\ \citealt{bro00}).
For standard IMFs (\eg\ Salpeter or Miller-Scalo), increasing the metallicity above solar can 
potentially give a slightly younger age. Experiments with the recent \citet{bru03} models show 
that the gross
SEDs of solar-metallicity models can be reproduced by models at 2.5 times solar metallicity with
ages reduced by about 30\%. However, any real stellar population will have a range of metallicities,
and chemically consistent evolutionary synthesis models (\eg\ \citealt{fri99,jim00}) show that 
populations with mass-weighted metallicities that are much higher than solar can still have
integrated fluxes dominated by stars with sub-solar metallicities in the rest-frame near-UV,
where line blanketing is important. This effect means that ages estimated from solar-metallicity
models are unlikely to be serious overestimates \citep{dun99}. In the case of KC68, 
there seems to be no way to avoid requiring almost all of the stars to be close to 1 Gyr old or
older. Furthermore, such solutions are about a factor of 3 less probable than the best fit, which is
for a very old (2--3 Gyr) population with little reddening. Thus, it is likely that the most
recent significant star formation in 4C\,23.56 KC68 occurred when the universe was $<1$ Gyr
old, although we cannot completely exclude somewhat later times. (Note that if we were to assume 
the fairly extreme solution of
a 1-Gyr-old population with internal reddening of $A_V\sim1.1$ mag, the star formation would 
have to be essentially complete by $z\sim3.8$, when the age of the universe would have been
1.6 Gyr).  The important point remains, in any case,
that there cannot have been significant recent star formation.

{\it Is the morphology really disk-like?}---As we have stated in \S\ref{aosect}, we cannot
exclude the possibility of a small $r^{1/4}$-law component, but any such component seems to be
insignificant in terms of the total luminosity of the system. The small projected axial ratio and the 
good fit to a pure exponential profile indicate that the system is strongly dominated by a disk.

In sum, the most straightforward and plausible interpretation remains that 4C\,23.56 KC68 is a 
well-settled disk of old stars at a redshift close to that of 4C\,23.56 itself. If this interpretation is 
indeed correct, what are the consequences? Firstly, something very close to the classical picture of
monolithic collapse must have occurred, with the important proviso that vigorous star formation
did not begin until after the gas had settled into a disk. Secondly, star formation must have proceeded at an extremely rapid rate in order to form the required $\sim3\times10^{11} M_{\odot}$ of stars over a time span that could not possibly be much more than $10^9$ years and that was probably
more like a few $10^8$ years. Thirdly, the star formation and processes associated with it must 
have been very efficient at processing and expelling the gas in order to have strongly quenched 
any subsequent significant star formation.  

In hierarchical formation models, including the standard CDM picture, galaxy morphologies are largely 
a reflection of their merger and accretion histories up to the epoch at which they are observed 
(\eg\ \citealt{ste03}). Semi-analytic models of galaxy formation (\eg\ \citealt{kau98}) suggest that
most massive spheroids underwent final assembly between redshifts of 1 and 2, and that most large
spiral disks were gradually acquired by smooth accretion of cold gas at redshifts $\sim1$ and lower,
after the last major merger \citep{kat91,nav94}. There appears to be little place in this scenario 
for a galaxy like 4C\,23.56 KC68, for at least two reasons: (1) the implied prodigious star
formation rate, which has to have been sustained for a few $10^8$ years {\it without} the help
of a major merging event (as indicated by
the lack of a detectable bulge), and (2) the building up of a massive galaxy apparently by smooth
accretion of cold gas, without any evidence for significant hierarchical merging.

We of course cannot say from this one example whether similar galaxies are common at high redshifts.
If they are, they would in one sense be the perfect progenitors for spheroidals: they comprise natural
reservoirs of stars formed over very short periods of time in deep potential wells at early epochs.
In short, their stars have all of the properties needed to explain the color-luminosity and
alpha-element-luminosity relations for spheroidals, provided only that the morphological transformation
from disks to spheroidals results from mergers of a small number of massive disk galaxies \citep{kau98}.
One obvious course for the future would be to explore the morphologies of galaxies selected in the 
same way as KC68 was with high resolution imaging, either with NICMOS on
{\it HST}, or with laser-guide-star AO on large ground-based telescopes.

\acknowledgments
We thank Jenny Patience for assisting with some of the Subaru observations,
and we thank the staff of the Subaru and Keck telescopes for their
excellent support for our observing programs. We have benefited from
discussions with Josh Barnes, John Kormendy, and Brent Tully.
This research has been partially supported by NSF grant AST03-07335. 
It made use of the NASA/IPAC Extragalactic Database (NED) 
which is operated by the Jet Propulsion Laboratory, California Institute of 
Technology, under contract with the National Aeronautics and Space 
Administration. The authors recognize the very significant
cultural role that the summit of Mauna Kea has within the indigenous
Hawaiian community and are grateful to have had the opportunity to
conduct observations from it.

\clearpage

\end{document}